# Reflectometry Measurements of the Loss Tangent in Silicon at Millimeter Wavelengths

Grace E. Chesmore[1], Tony Mroczkowski[2], Jeff McMahon[1], Shreya Sutariya[1,3], Alec Josaitis[1,4], and Leif Jensen[5]

[1]*University of Michigan Department of Physics*
*450 Church St, Ann Arbor, MI 48104 USA*
*Email: chesmore@umich.edu*

[2]*European Southern Observatory*
*Karl-Schwarzschild-Str. 2,*
*D-85748 Garching, Germany*
*Email: tony.mroczkowski@eso.org*

[3]*Wayne State University Department of Physics*
*42 W Warren Ave, Detroit, MI 48202 USA*

[4]*University of Cambridge Department of Physics*
*The Old Schools, Trinity Ln, Cambridge CB2 1TN, UK*

[5]*Topsil Semiconductor Materials A/S, Frederikssund, Denmark*

**ABSTRACT**

We report here on measurements of the reflectivity and loss tangent measured in the W-band (80-125 GHz) and D-band (125-180 GHz) in two samples of float zone silicon with intrinsic stoichiometry - one irradiated by neutrons, which increases the resistivity by introducing crystalline defects, and the other unperturbed. We find a $tan\,(\delta)$ of $2.8 \times 10^{-4}$ and $1.5 \times 10^{-5}$ for neutron-irradiated silicon and intrinsic silicon, respectively, both with an index of refraction of 3.41. The results demonstrate the applicability of silicon as a warm optical component in millimeter-wave receivers. For our measurements, we use a coherent reflectometer to measure the Fabry-Perot interference fringes of the reflected signal from dielectric slabs. The depth of the reflection nulls provides a sensitive measurement of dielectric losses. We describe the test setup which can also characterize scattering and transmission, and can provide detailed characterization of millimeter wave materials.

**INTRODUCTION**

Dielectric loss and its associated thermal emission is a key driver for sensitivity in millimeter wave receivers. A low loss implies low emission from a material, therefore limiting thermal noise propagating through to the detector. The contribution to the thermal noise due to loss, $T_{sys}$, can be calculated as $T_{sys}=T_{comp}(L-1)$, where $T_{comp}$ is the physical temperature of the component, $L = exp\,(2\,\pi\,n\,tan\,(\delta)\,t\,/\,\lambda_0)$ is the optical loss, $n$ is the index of refraction, $tan\,(\delta)$ is the loss tangent, $t$ is the sample thickness, and $\lambda_0$ is the free-space wavelength. The noise due to loss can be reduced in two ways: lowering the temperature of the component, and using materials with lower intrinsic loss.

Many millimeter wave experiments use cold optical components to minimize noise contributions due to this thermal emission. However, in many cases it is advantageous to include warm silicon components in optical designs. For example in the ALMA Band 1 receivers include a warm lens that also acts as the window into the cryostat. The original design uses a plastic (high density polyethylene) lens that contributes one-third of the overall receiver noise budget. Substitution of a lens with lower dielectric losses could therefore make a substantial impact on the sensitivity of this instrument.

Silicon is a potentially attractive material for such applications. Datta et al. 2013 showed that silicon can be machined to produce very high quality anti-reflective (AR) meta-surfaces to reduce the reflected power from the natural 30% to levels in the few tenths of a percent range [1]. The key development required is identification of a source of silicon with low dielectric losses at room temperature.

Since it was first reported in Parshin et al. 1995 [6] and included in the Lamb 1996 compendium of mm-wave material properties [5], the promise of silicon as an intrinsically low-loss mm/submm optical material has tantalized the astronomical community. More recently, Krupka et al. 2016 [4] reported on the result of a loss tangent in Ka-band and

Q-band (~27-40 GHz) on the order of $tan\,(\delta) \lesssim 10^{-5}$ at room temperature (~300 K) for samples of proton-irradiated high-purity FZ silicon, but measurements in W-band (~90 GHz) and for neutron-irradiated samples are lacking. We address this here using coherent reflectometry measurements.

**PROCEDURE**

**Optical Hardware**

A feed-horn source sends a millimeter wave signal to a parabolic mirror, which sends a plane wave toward the sample, typically at 10º angle of incidence. For these measurements the sample is a dielectric slab immersed in air. A portion of the signal is reflected the first air-dielectric boundary, and the remaining signal propagates through the sample, where it again partially reflects from the second dielectric-air boundary leading to a Fabry-Perot interference in the dielectric cavity defined by the sample. The reflected signal propagates to a second mirror followed by the receiver feed horn. The geometry is shown in Fig. 1.

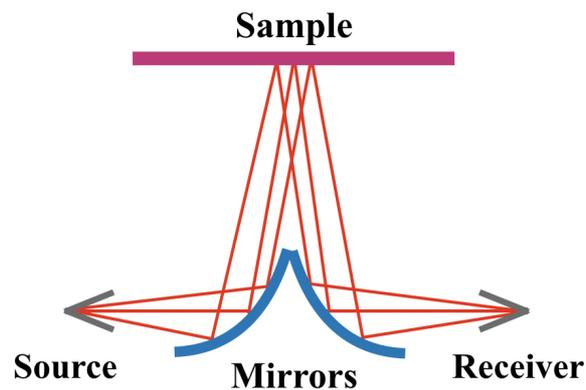

Fig. 1: Reflectometry setup. The red lines the beam as it travels from the source feed-horn, reflected off of the first mirror and to the sample. The beam then reflects off of the sample and to the second mirror, where it is reflected into the receiver.

Eccosorb is placed on surfaces outside the beam in order to reduce and absorb unwanted reflections. The alignment of the receiver is controlled with a three-axis stage. The tilt of the sample is controlled with a three-point micrometer mount. This angle is chosen by placing an aluminum plate in the sample holder and maximizing the received signal at a single frequency. Once the setup is aligned, a calibration data set is taken by measuring the reflectance of t aluminum plate as a function of frequency. This dataset serves to define perfect reflection. The aluminum plate is then replaced with the sample (silicon here) and the measurement is made. The response of the silicon sample measurement is divided by the response of the aluminum plate to yield a measurement of the reflectance of the sample as a function of frequency.

**Receiver Electronics**

A correlation receiver designed for a holography system is used for these measurements. The receiver compares a reference signal to a signal which has passed through the device, creating an interference pattern between the two. The holographic imaging setup is summarized in Fig. 2. The Re-configurable Open Architecture Computing Hardware (ROACH-2) board correlates the reference and modulated signals [7].

A millimeter-wave source sends a signal, ranging from 10.5-13 GHz, to a multiplier where passes through a passive multiplying chain. The multiplication factor varies with the source instrument, where the W-band (80-125 GHz) and D-band (125-180 GHz) millimeter-wave sources use multiplication factors of 9 and 12, respectively. The signal is modulated by the sample. The modulated and unmodulated (or "reference") signals are separately sent to two Pacific Millimeter harmonic mixers[1] (blue to yellow and blue to red boxes in Fig. 2. The mixers extract interference information caused by the offset frequency and send this information to the ROACH-2 FPGA, where the signals are correlated. This receiver outputs the amplitude and phase of the signal in narrow (50 MHz) spectral bands. Only the amplitude is used in this analysis, though we note that the phase information could be used to improve these measurements in the future.

---

[1] http://www.pacificmillimeter.com/HarmonicMixers.html

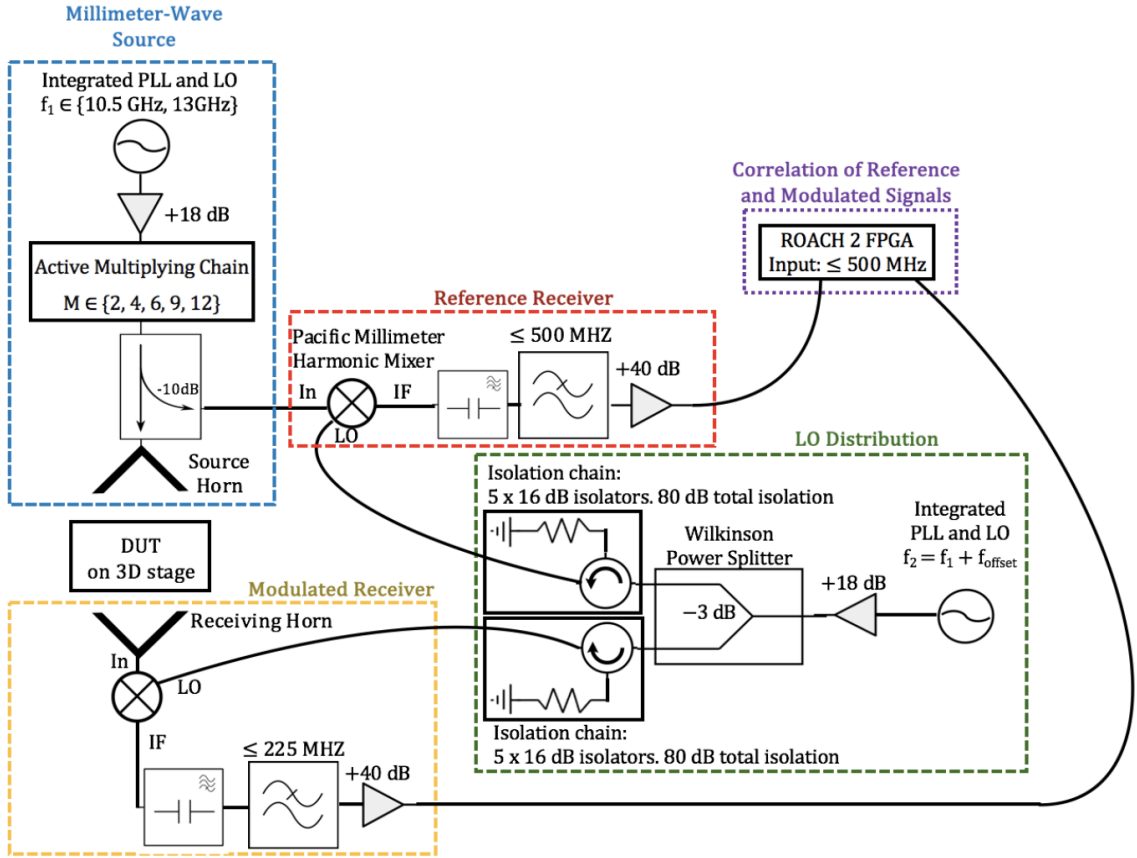

Fig. 2. Summary of the holographic imaging setup. A splitter sends one signal from a millimeter-wave source to a Pacific Millimeter harmonic mixer, while sending the same signal to be modulated by the device under test. The reference and modulated signal are mixed with an LO with an offset frequency $f_{offset}$ from the millimeter wave source. The two Pacific Millimeter harmonic mixers extract interference information caused by $f_{offset}$ and sends this to the ROACH-2 field programmable gate array (FPGA) board where the two signals are correlated.

**Samples: Neutron-irradiated and Intrinsic Silicon**

Two samples were provided by Topsil for testing. The samples as cut exhibited curvature and uneven thickness of approximately 15 mm. A professional third party was contracted to grind and polish the samples flat and parallel. The resulting thicknesses are reported in Table 1, and are accurate to scales $<< \lambda_0$ measured with a micrometer. The silicon is made with Float Zone mono growth in an oxygen-free environment, reducing risk of generating thermal donors during high temperature processes [2].

**REFLECTOMETRY AND TRANSMISSIVITY**

We measured the transmission and reflection of these samples. The data are shown in Fig. 3.

**Modeling**

We model the dielectric slab with a complex index $n_2 = n_r + in_i$. Air is treated with a purely real index $n_1 = 1$. The refracted angle in each layer *2* is calculated using Snell's law where $\theta_1$ is the incident angle and $n_1$ is the free-space index of the first layer (i.e. $n_1$ in (1)). The setup began at an incidence angle of 10°.

$$\theta_2 = arcsin\left(\frac{n_1}{n_2} sin\ \theta_1\right)$$

(1)

The experiment exclusively uses TM mode signals, where the reflected and transmitted coefficients theoretically predicted by [3]:

$$R = \frac{n_1 \cos\theta_2 - n_2 \cos\theta_1}{n_1 \cos\theta_2 + n_2 \cos\theta_1}$$

(2)

$$T = \frac{2n_1 \cos\theta_1}{n_2 \cos\theta_1 + n_1 \cos\theta_2}$$

(3)

$$\tan(\delta) = 2n_i/n_r$$

(4)

Using (2) and (3), where the parameter $n_2$ is the fitting parameter, the material index is extracted from measurements.

**Fitting**

The loss tangent (or equivalently complex part of the index) leads to a reduction in the depth of the reflection nulls for this signal. The cancellation of the incident and reflected waves is more perfect for lower loss tangent. Therefore the nulls in the reflection are crucial to get a reliable constraint on the loss. We also note that one of the dominant systematic effects is standing waves between the source and receiver horn. This systematic includes a reflection from the sample. Therefore this noise is minimized near the nulls. For this reason we fit the data only where the reflectance is below some threshold (typically R<0.01, though varying this by a factor of a few has no impact on the results). The weighing of each data point scales as the inverse of reflectivity to force the fit to focus on the depth of the nulls. We test that the analysis doesn't depend on this weighting by doing describe.

We implemented a Markov Chain Monte Carlo algorithm for fitting the reflectivity model to our measurements using the Python package pyMC, with the real and imaginary components of the index of refraction as fitting parameters [8]. The Markov Chain Monte Carlo algorithm samples the probability distribution of each parameter, in this case the probability of two parameters in the reflectivity model matching the measured reflectivity of the sample. This method was used to fit the measurement to the model, yielding, the real and imaginary components of index of refraction of the sample. The loss, $\tan(\delta)$, of the material is then calculated using (4).

The error bars of loss values are calculated with two methods. The first method is the MCMC error computation with the pyMC package in Python. To check these values, each null is fit individually using the MCMC algorithm, yielding an array of parameter values. The root-mean-square (RMS) of these values is then calculated, yielding the RMS error. Both methods yielded nearly equivalent loss error results, and we therefore report the MCMC loss error in Table 1. The error in index of refraction is not reported as it is likely included in the uncertainty of the thickness, while the MCMC reported a negligible error for this parameter.

**Results**

The index of refraction of neutron and intrinsic silicon are nearly identical with $n$ =3.415 for neutron-irradiated silicon, and $n$ =3.412 for the intrinsic silicon samples. The neutron-irradiated sample has $\tan(\delta) = 2.8 \times 10^{-4}$, while the intrinsic silicon is nearly an order of magnitude better with $\tan(\delta) = 1.5 \times 10^{-5}$. See Table 1 for a summary of the measured and inferred properties.

The reason for the higher loss in the neutron-irradiated sample is unclear, as the loss tangent should improve for materials with higher resistivity if dominated by the conduction loss term. As the resistivities of the neutron-irradiated and unmodified intrinsic silicon samples were measured by Topsil to be >100 k$\Omega$-cm and >50 k$\Omega$-cm, respectively, the expectation is that the neutron-irradiated sample should have less loss. One possible explanation is that the lattice defects introduced by neutron irradiation contribute to the loss.

The resulting values for $\tan(\delta)$ from fits to our measurements of the unmodified intrinsic silicon sample are comparable to those reported in Parshin et al. 1995 [6]. Further, our measured low loss in intrinsic silicon demonstrates it can be a compelling material for use in warm optics. Krupka et al. 2016 [4] also report a $\tan(\delta) \cong 1.2 \times 10^{-5}$ for proton-irradiated silicon, which is lower by more than 3-$\sigma$ than our measurements of intrinsic silicon, but on the same order of magnitude.

Our work shows promise for the use of intrinsic silicon for room-temperature optical components, such as those required for the lower bands of the Atacama Large Millimeter/Sub-millimeter Array (ALMA) and potential future CMB experiments.

Table 1. Optical properties of silicon. Properties extracted from the data when fit with the theoretical model. $n$ is the index of refraction, D is the thickness of the sample, and $tan(\delta)$ is the loss tangent.

|  | Silicon | |
| --- | --- | --- |
| Type: | Neutron | Intrinsic |
| $n$ | 3.415 | 3.412 |
| $tan(\delta) \times 10^{-5}$ | 28.93 ± 1.21 | 1.47 ± 0.09 |
| D [mm] | 13.68 ± 0.1 | 13.56 ± 0.1 |

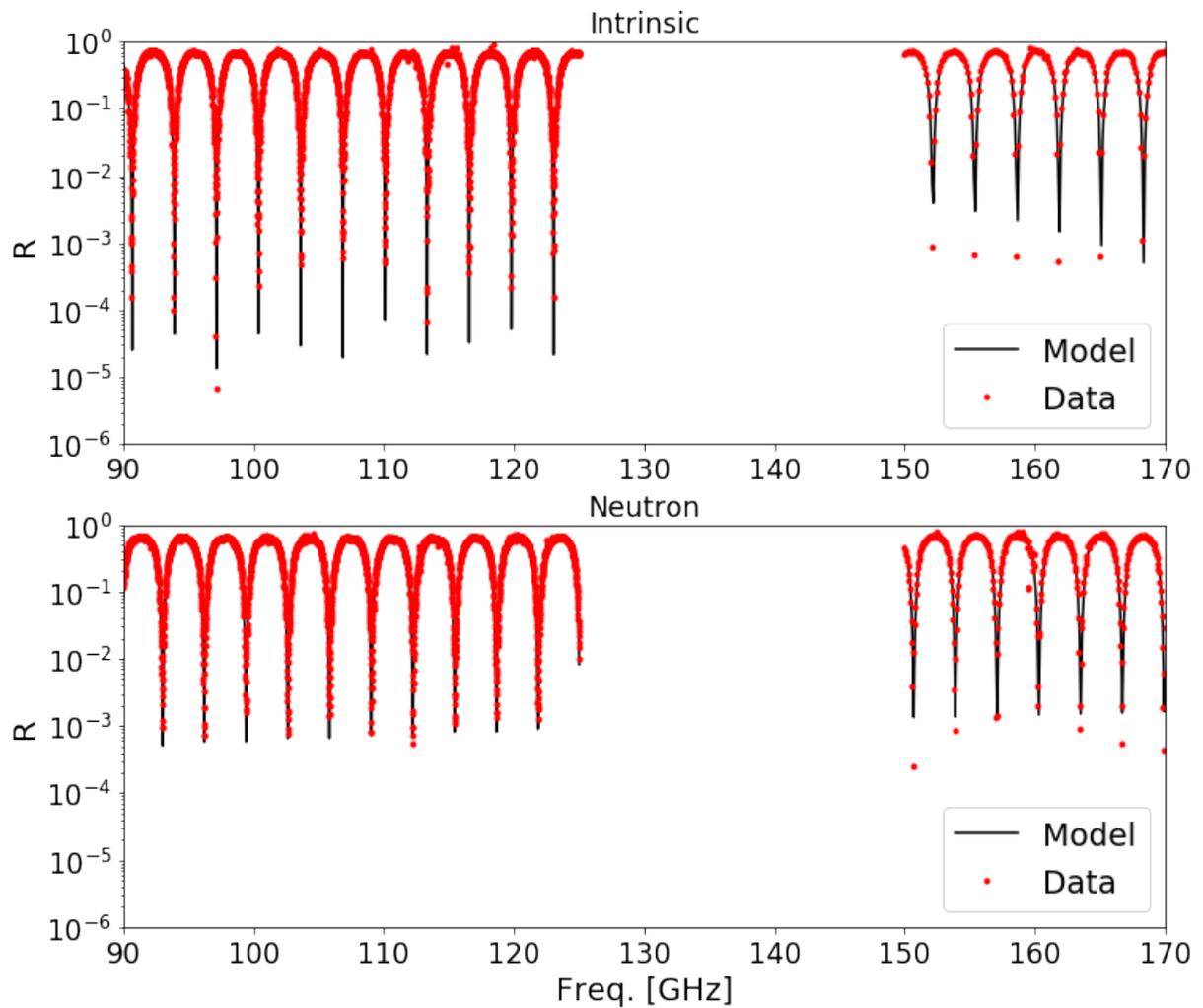

Fig. 3. Reflectivity of intrinsic (top) and neutron-irradiated (bottom) silicon against frequency in GHz.